# Middle Infrared and THz Sources at AREAL

V. V. Sahakyan[*], A. A. Sargsyan, G. S. Zanyan, T. S. Vardanyan, B. A. Grigoryan,
V. M. Tsakanov
CANDLE SRI, 31 Acharyan, 0040 Yerevan, Armenia

**Abstract:** AREAL (Advanced Research Electron Accelerator Laboratory) is an electron linear accelerator with a laser driven RF gun constructed at CANDLE Synchrotron Research Institute. After the successful operation of the gun section at 5 MeV, a program of facility energy growth up to 50 MeV has been launched, which includes the construction of two advanced experimental stations for the generation of middle infrared and THz radiation. This paper presents the design of these stations and gives the characteristics of expected radiation based on numerical and theoretical studies.

## 1. Introduction

The THz range of electromagnetic spectrum has for a long time been considered as a rather poorly explored region since its first observation about one hundred years ago. However, during the past few decades this situation has changed with the rapid development of coherent THz sources, such as solid state oscillators, gas and quantum cascade lasers, laser driven THz emitters [1].

The advent of laser-based coherent THz emission has radically changed linear spectroscopy and imaging and helped to significantly improve our fundamental understanding of matter. THz pulses have since been used as sensitive probes of fundamental low-frequency excitations in solids, liquids or gases. Combined with synchronized femtosecond pulses in the near-infrared, visible, ultraviolet, or even x-ray range, THz pulses are ideal tools for studying the dynamic response of matter to a strong field, varying rapidly in time but still much slower than the field of visible laser light. Thus, THz frequency is a frontier region for a research in physics, chemistry, biology, materials science and medicine [2].

The promising sources of THz radiation are accelerator-based sources. Numerous accelerator-based THz sources are operational worldwide as user facilities providing high radiation power and brightness. Such THz sources produce radiation via one of the following principles [3]: undulator radiation, dipole radiation from a bending magnet, transition radiation, diffraction radiation, edge radiation at bending magnets or undulators, Cherenkov radiation and Smith–Purcell radiation. These techniques require either ultrashort electron bunches or electron bunches with longitudinal density modulations on a THz wavelengths scale. In storage rings the electron bunch durations are relatively long, however, some of them can operate in a special mode, called low alpha mode, which allows shortening the electron bunches to the required few picosecond level.

Although coherent THz radiation was first observed at storage rings, however, the true potential of accelerator-based THz emitters was revealed once it was observed at linacs, where electron bunches are typically much shorter than those of a usual storage rings.

The electron bunches generated in AREAL linear accelerator [4], which is under operation at CANDLE research institute, satisfy the requirements for production of high power coherent THz radiation. The current paper is devoted to the study and development of two advanced experimental stations based on AREAL for the generation of coherent radiation in middle infrared and THz spectrum. The paper is composed of the following sections: in section 2 the current status of AREAL linear accelerator and its upcoming 50 MeV upgrade program is described. In section 3 a detailed characterization of dedicated radiation stations is given. The results of performed numerical simulations and theoretical calculations are presented. Finally, the summary is given in section 4.

## 2. AREAL 50 MeV Upgrade

Photo cathode RF gun based AREAL linac is under operation at CANDLE Synchrotron Research Institute. The aim of this new facility is to generate high brightness electromagnetic radiation with sub-picosecond pulse duration for advanced experimental studies in the fields of accelerator technology and dynamics of ultrafast processes. The construction and commissioning of RF photo gun and diagnostic stations were completed in May 2014. After a successful operation of the gun section, AREAL facility upgrade program has been launched. After the gun section two 1.6 m long S-Band traveling wave accelerating structures, with maximum gradient of 15 MV/m, will be installed. Two quadrupole doublets and one triplet combined with corrector magnets will be used for beam focusing and trajectory control. After the second quadrupole doublet a 60° bending dipole magnet will be placed allowing to switch between two beamlines. This magnet will also be used for electron beam energy, energy spread and bunch length measurements. In addition, two other bending magnets will be located after undulator sections for electron and photon beam separation. The schematic layout of AREAL linac after the upgrade is shown in Fig. 1 [5].

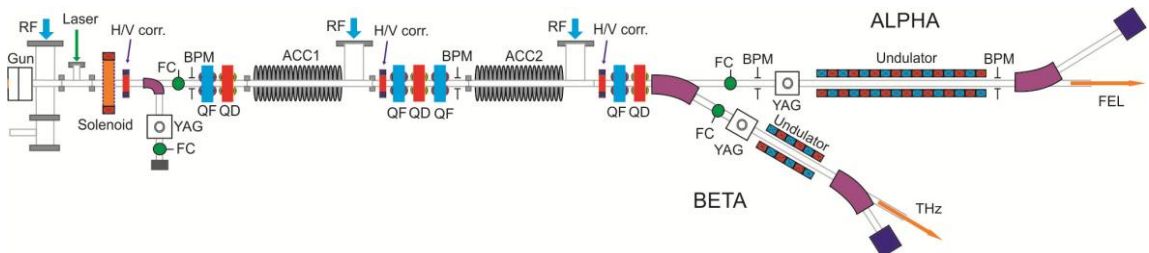

Figure 1: Schematic layout of AREAL linear accelerator.

The expected beam main parameters of AREAL linac are listed in Table 1.

Table 1: Beam parameter list

| Energy | ≤ 50 MeV |
|---|---|
| Bunch charge | 10-250 pC |
| Transv. norm. emit. | < 0.5 mm.mrad |
| Bunch length (rms) | 0.4 – 9 ps |
| RMS energy spread | < 0.15% |
| Number of bunches per pulse | 1-16 |
| Repetition rate | 1-50 Hz |

The upgrade program presumes an increase of beam energy up to 50 MeV and exploitation of ALPHA (Amplified Light Pulse for High-end Applications) and BETA (Booster for Emerging Technology Accelerators) experimental stations. The aim of these stations is to create middle infrared and THz coherent radiation sources using AREAL electron beam, as well as, provide bases for experimental investigations in the fields of advanced accelerator concepts and new radiation sources [4].

According to the current schedule the upgrade program will be completed in 2018.

### 3. Middle Infrared and THz Stations

Currently, two advanced accelerator-based techniques for coherent middle infrared and THz radiation generation are under consideration: the principles of SASE FEL [6] and superradiant enhancement of radiation from short electron bunches [7].

### 3.1 SASE FEL Radiation on AREAL

The successful generation of SASE FEL implies a small electron beam natural emittance and energy spread, in order to match with diffraction-limited photon beam phase-space and to avoid radiation spectrum broadening. AREAL electron beam parameters enable an effective SASE FEL generation of tens of MW power in an appropriate undulator having a length of several meters.

For this purpose, a 4.5 m long planar fixed gap permanent magnet undulator, based on NdFeB material, will be used [8]. Table 2 gives an overview over its key parameters.

Table 2: Undulator parameters

| Period length (mm) | 27.3 |
|---|---|
| Gap (mm) | 12 |
| Peak Field (T) | 0.468 |
| K parameter | 1.17 |
| Number of poles | 327 |

We have considered two energy cases: 50 MeV (maximum design energy) and 30 MeV, which is the lowest energy allowing the effective generation of SASE process using AREAL electron beam. According to the well-known formula of planar undulator radiation wavelength:

$$\lambda = \frac{\lambda_u}{2\gamma^2}\left(1 + \frac{K^2}{2}\right) \qquad (1)$$

where $\lambda_u$ is undulator period length, $\gamma$ is the Lorentz factor and $K$ is undulator parameter, by varying electron beam energy in the range of 30 – 50 MeV, the radiation wavelength range of 2.4 – 6.7 µm will be covered, which lies in the middle infrared part of electromagnetic spectrum.

The SASE FEL performance has numerically been studied using GENESIS 3D time-dependent simulation code [9]. In simulations the electron beam and undulator parameters listed in Tables 1, 2 have been used and 20 initial seeds for the random number generator for particle phase fluctuation (shot noise) have been considered.

In Fig. 2 the radiation power along undulator for different energy cases is shown. As it is expected with the growth of electron beam energy the FEL saturation length increases, however it is still less than the undulator length. The saturation length for 50 MeV energy case will be around 3.2 m. According to GENESIS simulations, the saturation power of 40 – 60 MW and the pulse energy of 60 – 100 µJ can be achieved.

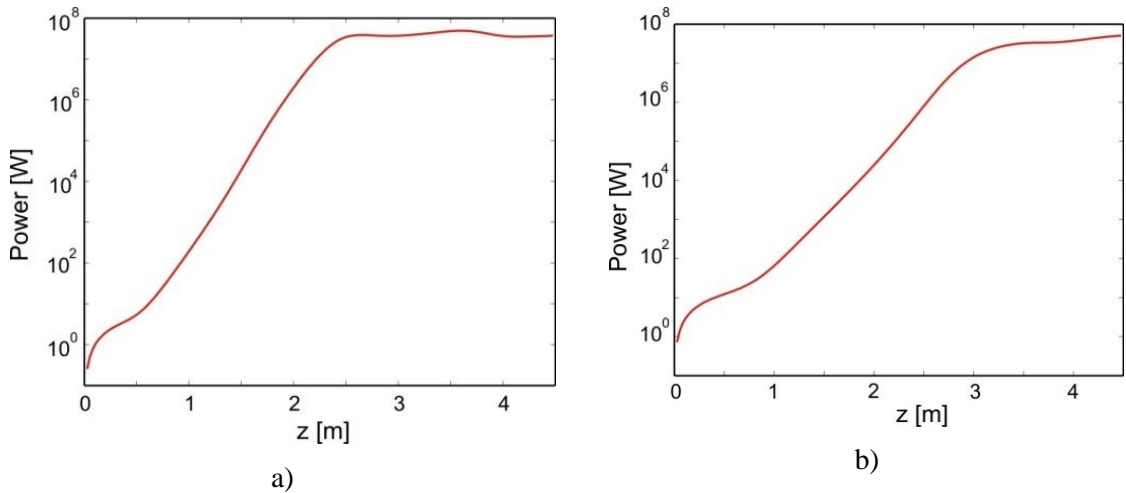

Figure 2: Radiation energy along the undulator for 50 MeV (a) and 30 MeV (b) beam energy cases.

The power distribution along the bunch at saturation point for the considered 30 MeV and 50 MeV beam energy cases are presented in Fig. 3.

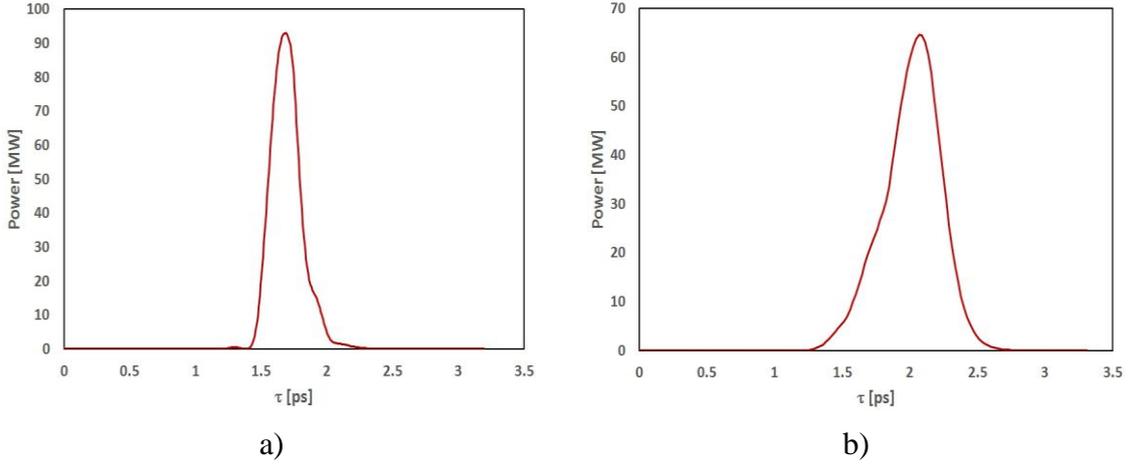

Figure 3: Power distribution along the bunch for 50 MeV (a) and 30 MeV (b) beam energy cases.

### 3.2 Superradiant Undulator Radiation on AREAL

Superradiant emission from electron bunches becomes significant for frequencies sufficiently lower than the inverse of the bunch duration [10]. In general, for a certain frequency $\nu$, the emitted pulse energy can be derived from the emission characteristics of the radiator for a single electron $W_e$ as follows:

$$W_b = N_e W_e [1 + (N_e - 1) f(\nu)],$$

$$W_e = 2 \frac{\pi q_e^2 N_p}{3 \varepsilon_0 \lambda_u} K^2 \gamma^2, \qquad (2)$$

where $N_e$ is the number of electrons, $q_e$ is the electron charge, $N_p$ is the number of undulator poles, $\varepsilon_0$ is the vacuum permeability and

$$f(\nu) = \exp\left[-2(\pi\nu)^2 \left(\frac{\tau}{2.35}\right)^2\right], \qquad (3)$$

is the so-called form factor (here given for a Gaussian-shaped bunch of FWHM - duration $\tau$) which quantifies the degree of superradiance. It is a dimensionless, frequency-dependent parameter that describes the fraction of the electron bunch that satisfies the superradiance condition. When the electron bunch length is much shorter than the radiation wavelength, the form factor is close to 1 and the undulator radiation energy is proportional to the square of the number of electrons per bunch.

The form factors for Gaussian-shaped bunches with 40 fs, 400 fs and 4000 fs FWHM-duration are shown in Fig. 4 a). As can be seen from the plot for a 40 fs (FWHM) Gaussian-shaped bunch, the emission is practically fully superradiant from the whole electron bunch for frequencies up to 3 THz. For a 400 fs bunch, a substantial emission at frequencies up to 1 THz can still be expected. However, in case of 4 ps long bunches, for frequencies from 0.3 THz to 3 THz, the emission coherency is practically lost.

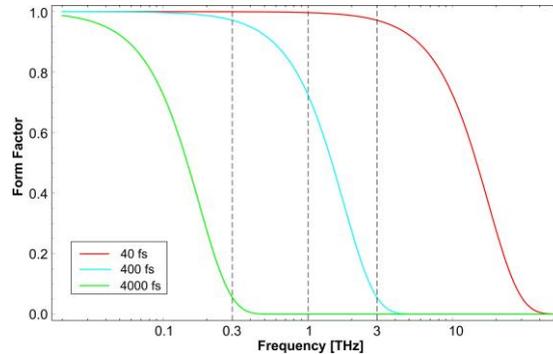

Figure 4: The form factor for different bunch durations

In the current design of AREAL linac the shortest achievable bunch length is 400 fs. It is also planned to add a new module for laser pulse compression in laser system, which will allow to produce even shorter bunches (bellow 100 fs). Taking into account the aforementioned capacity of AREAL linac, it is foreseen to have the second radiation source at BETA beamline (see Fig. 1), based on superradiant undulator radiation principle. In order to extend the capabilities of AREAL linac as a user facility, the undulator system for superradiant radiation is planned to be a helical fixed gap device for the production of elliptically polarized radiation. Note that the predefined operation of this source requires the compensation of dispersion induced by dipole magnet. Thus, a dispersion compression scheme should be added at BETA beamline, which is under consideration.

The undulator system is currently under development. Because of available space restrictions at BETA beamline the total length of the undulator should not exceed 2.5 m. With the current design this undulator is assumed to have a period length of 12 cm with the K value equal to 1.2. By using this undulator, the frequency range from 3 to 10 THz with coherent radiation energy up to 4 mJ can be covered by varying electron beam energy from 20 to 35 MeV.

## 4. Summary and Outlook

This paper presents the overview of middle infrared and THz radiation sources based on AREAL linac. The 50 MeV upgrade program of AREAL linac and approaches for the realization of two advanced experimental stations are described. One of them will operate based on SASE FEL principle and produce radiation in the middle infrared range of electromagnetic spectrum. GENESIS time-dependent simulation results for the expected radiation parameters are presented. The second station will be based on superradiant enhancement of radiation from short electron bunches. It was shown, that with the implementation of 2.5 m long helical undulator, elliptical polarized radiation of energy up to 4 mJ can be generated in the range from 3 to 10 THz, using bunches with 100 fs duration. The production of such electron bunches will become possible after upcoming modifications of the existing laser system.

Note that the photon beam transport and diagnostic systems are currently under development.

**Acknowledgments**

This work is supported by the RA MES State Committee of Science, in the frame of the research project № 16YR-1C009 and was made possible in part by a research grant from the Armenian National Science and Education Fund (ANSEF) based in New York, USA.

**References**

[1] G.P. Gallerano et al., *Overview of Terahertz Radiation Sources*, Proceedings of FEL'04, Trieste, Italy (2004), FRBIS02.

[2] B. Zhu, Y. Chen et al., *Terahertz Science and Technology and Applications,* PIERS Proceedings, Beijing, China (2009).

[3] Anke-Susanne Müller, *Accelerator-Based Sources of Infrared and Terahertz Radiation*, Rev. Accl. Sci. Tech., 03 (2010), 165.

[4] N. Stojanovic, M. Drescher, *Accelerator- and laser-based sources of high-field terahertz pulses*, J. Phys. B: At. Mol. Opt. Phys. 46 (2013) 192001.

[5] V. Tsakanov, G. Amatuni, A. Sargsyan, G. Zanyan, V. Sahakyan et al., *AREAL Test Facility for Advanced Accelerator and Radiation Source Concepts*, Nucl. Instrum. and Meth. A 829 (2016) 284.

[6] A. A. Sargsyan et al., *An Overview of Beam Diagnostic and Control Systems for 50 MeV AREAL Linac*, 2017, JINST 12 T03004

[7] E. L. Saldin, E.A. Schneidmiller, M.V. Yurkov, *The Physics of Free Electron Lasers*, Springer, Berlin, Heidelberg (2000).

[8] G. P. Williams, *High–power terahertz synchrotron sources*, Phil. Trans. R. Soc. Lond. A 362 (2004), 403.

[9] J. Pflüger et al., *Magnetic Characterization of the Undulator for the VUV-FEL at the TESLA Test Facility*, TESLA FEL-Report 1999-07

[10] S. Reiche, *GENESIS 1.3: a fully 3D time-dependent FEL simulation code*, Nucl. Instr. Meth. In Phys. Res. A 429 (1999) 243-8.

[11] J. D. Jackson, Classical Electrodynamics, 3rd Edition, pp. 832. ISBN 0-471-30932-X. Wiley-VCH , 1998.